\documentclass[aps,prl,twocolumn,floatfix, showpacs]{revtex4-1}
\usepackage{graphicx}
\usepackage{dcolumn}
\usepackage{bm}
\usepackage{epstopdf}

\usepackage{array}
\usepackage{subfigure}

\newcommand{\be}{\begin{equation}}
\newcommand{\ee}{\end{equation}}

\newcommand\Ro{\mbox{\textrm{Ro}}}  
\newcommand\Pra{\mbox{\textrm{Pr}}}  
\newcommand\Ra{\mbox{\textrm{Ra}}}  
\newcommand\Nu{\mbox{\textrm{Nu}}}  

\def\gtwid{\mathrel{\raise.3ex\hbox{$>$\kern-.75em\lower1ex\hbox{$\sim$}}}}
\def\ltwid{\mathrel{\raise.3ex\hbox{$<$\kern-.75em\lower1ex\hbox{$\sim$}}}}

\begin{document}

\title{Finite-size effects lead to supercritical bifurcations in turbulent rotating Rayleigh-B\'enard convection}

\author{Stephan Weiss$^1$}
\author{Richard J.A.M. Stevens$^2$}
\author{Jin-Qiang Zhong$^1$}
\author{Herman J.H. Clercx$^{3,4}$}
\author{Detlef Lohse$^2$}
\author{Guenter Ahlers$^1$}

\affiliation{$^1$Department of Physics, University of California, Santa Barbara, CA 93106, USA}
\affiliation{$^2$Department of Science and Technology and J.M. Burgers Center for Fluid Dynamics, University of Twente, P.O Box 217, 7500 AE Enschede, The Netherlands}
\affiliation{$^3$Department of Applied Mathematics, University of Twente, P.O Box 217, 7500 AE Enschede, The Netherlands}
\affiliation{$^4$Department of Physics and J.M. Burgers Centre for Fluid Dynamics, Eindhoven University of Technology, P.O. Box 513, 5600 MB Eindhoven, The Netherlands}

\date{\today}

\begin{abstract}
In turbulent thermal convection in
 cylindrical samples of aspect ratio $\Gamma \equiv D/L$ ($D$ is the diameter and $L$ the height)
 the Nusselt number \Nu\ is enhanced when the sample is rotated about its vertical axis, because of the formation of Ekman vortices that extract additional fluid out of thermal boundary layers at the top and bottom. We show from experiments and direct numerical simulations that the enhancement occurs only above a bifurcation point at a critical inverse Rossby number $1/\Ro_c$, with $1/\Ro_c \propto 1/\Gamma$. We present a Ginzburg-Landau like model that explains the existence of a bifurcation at finite $1/\Ro_c$ as a finite-size effect. The model yields the proportionality between $1/\Ro_c$ and $1/\Gamma$ and is consistent with several other measured or computed system properties.
\end{abstract}

\pacs{47.27.te,47.32.Ef,47.20.Bp,47.27.ek}

\maketitle

Turbulence, by virtue of its vigorous fluctuations,  is expected to sample all of phase space over wide parameter ranges.  This viewpoint implies that there should not be any 
 bifurcations between different turbulent states. Contrary to this, several cases of discontinuous transitions have been observed recently in turbulent systems \cite{rav04,mon07,rav08}. When they occur, they are likely to be provoked either by changes in boundary conditions or boundary-layer structures, or by discontinuous changes in the  large-scale structures,  as a parameter is varied.

\begin{figure}
\includegraphics[width=0.4\textwidth]{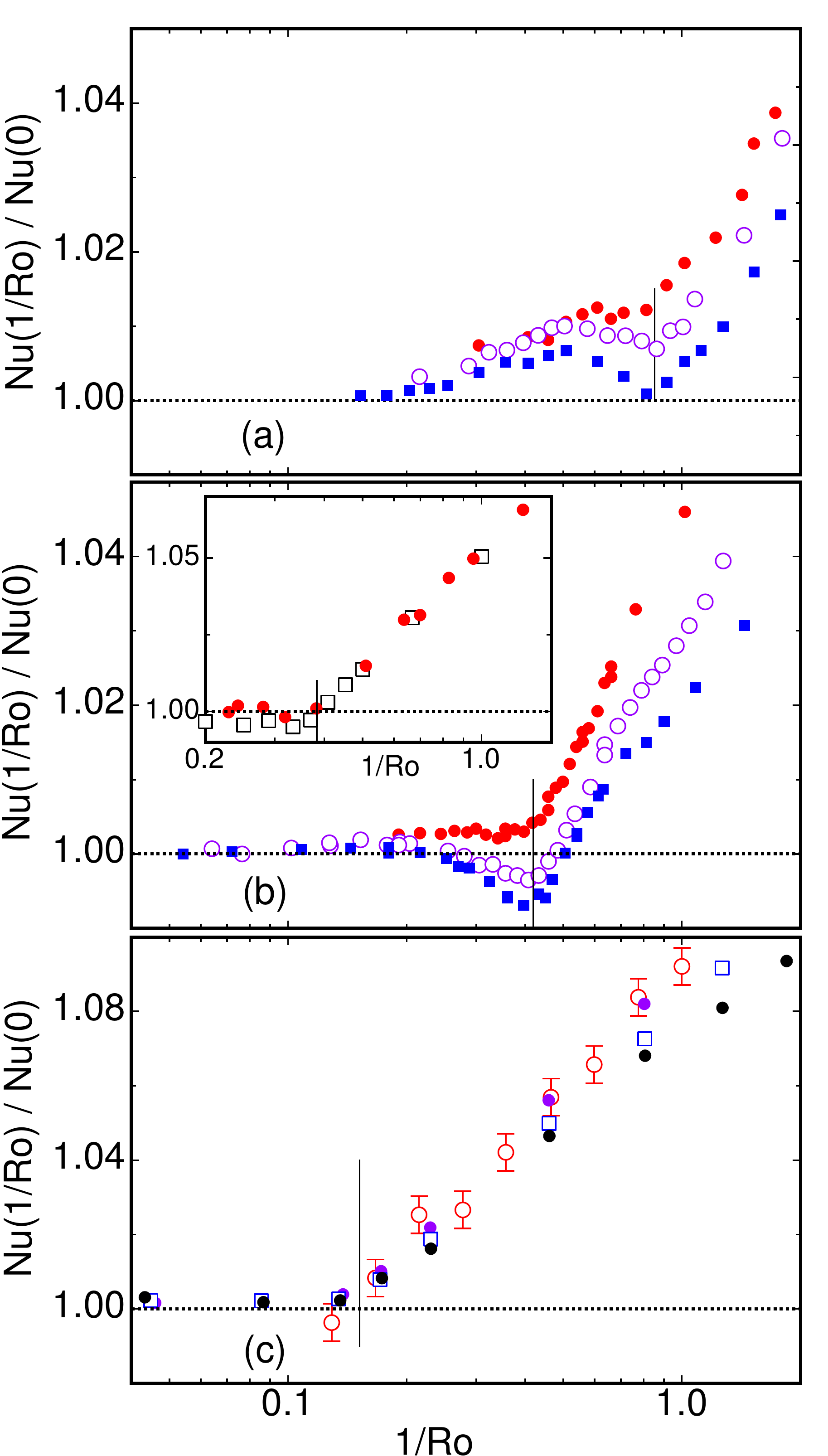}
\caption{
(Color online) The Nusselt number $\Nu(1/\Ro)$, normalized by $\Nu(0)$ without rotation, as a function of the inverse Rossby number $1/\Ro$. Data are from experiments unless mentioned otherwise. 
(a): $\Gamma = 0.50$ and $\Pra = 4.38$; Rayleigh numbers $\Ra = 9.0\times 10^9$ (solid circles, red online), $1.8\times 10^{10}$ (open circles, purple online), and $3.6\times 10^{10}$ (solid squares, blue online). 
(b): $\Gamma = 1.00$. Main figure: $\Pra = 4.38$, $\Ra = 2.25\times 10^9$ (solid circles, red online), $8.97\times 10^9$ (open circles, purple online), and $1.79\times 10^{10}$ (solid squares, blue online). 
Inset: $\Pra = 6.26$ and $\Ra = 2.73\times 10^8$. 
DNS: open squares (black online). 
Experiment: solid circles (red online). 
(c): $\Gamma = 2.00$ and $\Pra = 4.38$; the data are for $\Ra = 2.91\times10^8$ (open circles, red online, DNS; solid circles, purple online, experiment), $5.80\times10^8$ (open squares, blue online), and $1.16\times10^9$ (solid squares, black online). Note the different vertical scale for (c) compared to (a) and (b). The small vertical lines indicate the locations of the bifurcation points.} 
\label{fig:Nu}
\end{figure}

Recently some of us \cite{ZSCVLA09,SZCAL09} reported on the effect of rotation about a vertical axis at a rate $\Omega$ on turbulent convection in a fluid heated from below and cooled from above
(known as Rayleigh-B\'enard convection or RBC; for recent reviews, see \cite{AGL09,LX10,Ah09}). For a cylindrical sample of aspect ratio $\Gamma \equiv D/L = 1.00$ ($D$ is the diameter and $L$ the height) a supercritical bifurcation was found, both from experiments and from direct numerical simulation (DNS) of the Boussinesq equations of motion. At a {\it finite} $\Omega$, as expressed by the inverse Rossby number $1/\Ro \propto \Omega$ (to be defined explicitly below),  there was a {\it sharp} transition from a state of nearly rotation-independent heat transport (as expressed by the Nusselt number \Nu\ to be defined explicitly below) to one in which \Nu\ was enhanced by an amount $\delta \Nu(1/\Ro)$. This is illustrated by the data shown in Fig.~\ref{fig:Nu}. The increase of \Nu\ was attributed to Ekman pumping 
\cite{Ro69,ZES93,VE02b,KCG06,ZSCVLA09,SZCAL09,SCL10b, KSNHA09,NBS10}, i.e.  to the 
 formation of (cyclonic) vertical vortex tubes (``Ekman vortices''),  
which extract and vertically transport additional fluid from the boundary layers (BLs) and thereby enhance the heat transport. The bifurcation was located at a critical value $1/\Ro_c \simeq 0.40$ \cite{SZCAL09}. The reason for the existence of the bifurcation at $1/\Ro_c > 0$ hitherto had  
 not been understood. 
While such bifurcations are common near the onset of RBC in the domain of pattern formation
\cite{BPA00}, their existence in the turbulent regime implies a paradigm shift. 

In this Letter we report on further experiments for samples with $\Gamma = 2.00$, 1.00,  and 0.50
 which (i) all show bifurcations between different turbulent states and (ii) reveal
 that $1/\Ro_c$ varies approximately in proportion to $1/\Gamma$. We offer an explanation of these and other phenomena in terms of a phenomenological Ginzburg-Landau like description which predicts a finite-size effect upon the vortex density $A$. 
 
 We assumed that the relative Nusselt enhancement $\delta \Nu(1/\Ro)/\Nu(0)$ is proportional to the average $\bar A$ of $A$ over a horizontal cross section of the sample near the BLs.  Consistent with the DNS that we report here, we assumed that $A$ vanishes at the sample side wall. For the infinite system the model predicts that $A$, and thus $\delta \Nu/\Nu(0)$, increases linearly from zero starting at $1/\Ro = 0$.
 For the finite system the model gives a threshold shift proportional to $1/\Gamma$ as found in the experiment and by DNS. The shift is predicted to be followed by a linear increase of $\bar A$ in proportion to $(1/\Ro) - (1/\Ro_c)$ which yields $\delta \Nu/\Nu(0) = S_1(\Gamma) (1/\Ro - 1/\Ro_c)$. The model gives an initial slope $S_1(\Gamma)$ that decrease with decreasing $\Gamma$, again consistent with DNS and measurements. From DNS we show that $A$ decreases to zero near the side wall over a length that is consistent with an estimate of a healing-length $\xi$ based on the model.  Thus, we found consistency between the model predictions and all properties that we were able to either measure or compute from DNS.

Before proceeding, we define the relevant dimensionless parameters. The inverse Rossby number is given by
$1/\Ro=  (2\Omega)/(\sqrt{\beta g \Delta T/L})$,
were $\Omega$ is the rotation rate in rad/s, $\beta$ the isobaric thermal expansion coefficient, $\Delta T$ the temperature difference between the bottom and top plate, and $g$ the gravitational acceleration.
The Rayleigh number is $\Ra = (\beta g \Delta T L^3)/(\nu\kappa)$,
were $\kappa$ and $\nu$ are the thermal diffusivity and the kinematic viscosity respectively. The Nusselt number is given by $\Nu = (Q L)/(\Delta T \lambda)$
where $Q$ is the heat-current density and $\lambda$ is the thermal conductivity. Finally, the Prandtl number is $\Pra = \nu/\kappa$.  

In Fig.~\ref{fig:Nu} we show experimental 
and numerical data 
{\footnote{In Ref.~\cite{ZSCVLA09} we described the numerical method -- a finite difference solver for the Boussinesq equations
with the Coriolis force added to account for the rotation. E.g., the simulations at $Ra=2.91\times10^8$, $Pr=4.38$, and $\Gamma = 2$
 were performed on a $769 \times 385 \times 289$ grid in the azimuthal, radial, and axial direction respectively, which yielded sufficient resolution in the BLs and the 
 bulk \cite{SVL10}. Special care was taken for the azimuthal and radial resolution in order to keep the flow well resolved close to the sidewall boundaries in this large box.}
  for $\Nu(1/\Ro)/\Nu(0)$ as a function of $1/\Ro$ for several values of \Ra.
From top to bottom the three panels are for $\Gamma = 0.50$, 1.00, and 2.00,
 respectively 
 {\footnote{Some of the data for $\Gamma = 1.00$ had been published 
before \cite{ZSCVLA09,SZCAL09}.}.
One sees that  there is considerable structure even below the bifurcation, particularly at the larger \Ra.
To our knowledge the origin of this structure is not known in detail.  One sees that there are  clear breaks in the curves, {\it e.g.} for $\Gamma = 1.00$ (Fig.~\ref{fig:Nu}b) at $1/\Ro \simeq 0.4$, indicating the bifurcation to a different state. The location of this transition is within our resolution independent of \Ra. 

\begin{figure}
\includegraphics[width=0.4\textwidth]{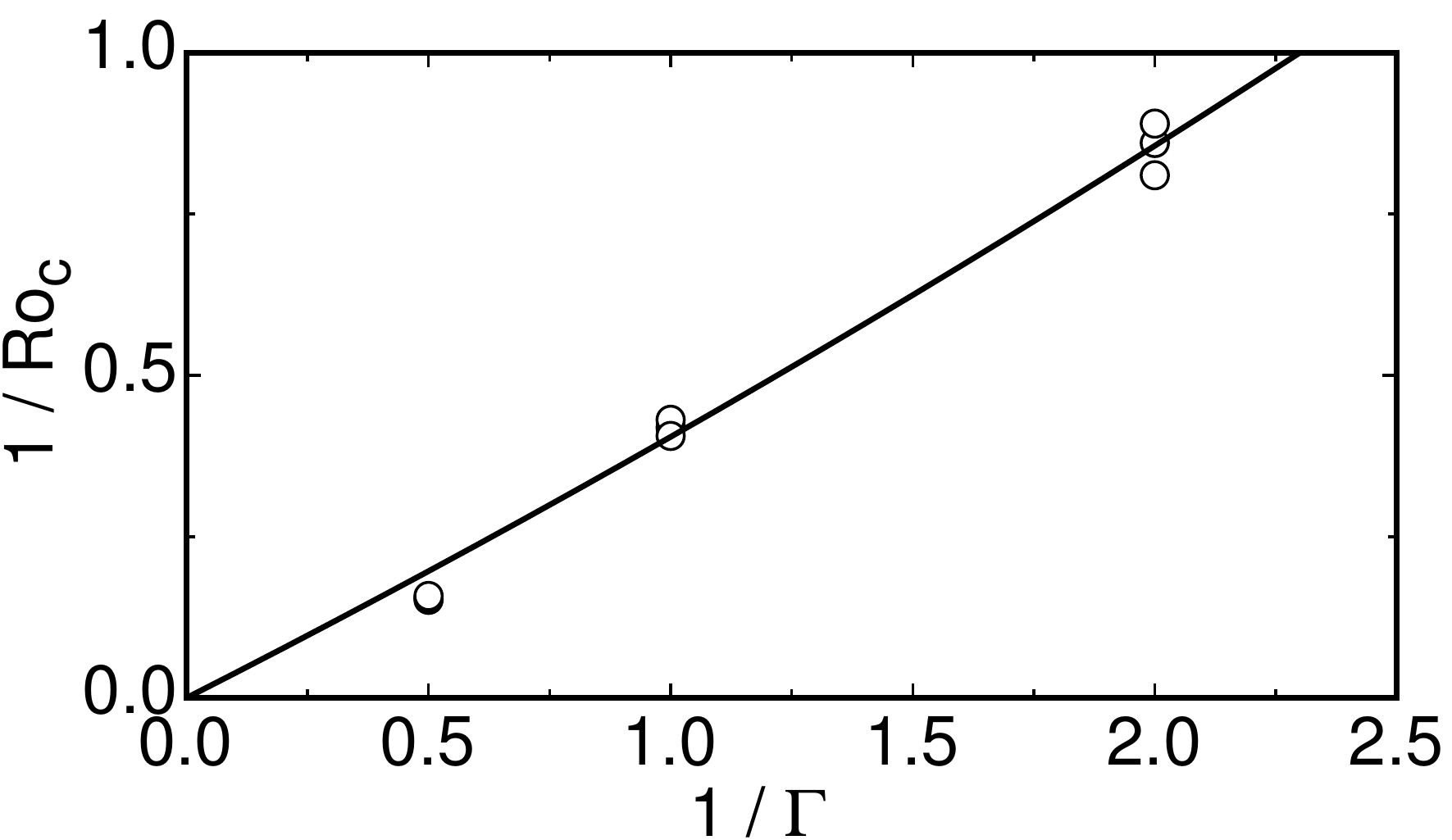}
\caption{The critical inverse Rossby number $1/\Ro_c$ as a function of the inverse aspect ratio $1/\Gamma$ for $\Pra = 4.38$ and different $\Ra$ (see Fig.~\ref{fig:Nu}). The line is a fit of a quadratic equation with the constant coefficient set to zero.}
\label{fig:Ro_c}
\end{figure}

In Fig. \ref{fig:Ro_c} we plotted all available data for $1/\Ro_c$ for $\Pra = 4.38$ (and different $\Ra$) as a function of $1/\Gamma$. The line shown there is a fit of 
\be
\frac{1}{\Ro_c} = \frac {a}{\Gamma}\Big (1+\frac{b}{\Gamma}\Big )
\label{eq:polynomial}
\ee
to the data. Its coefficients are $a = 0.381$ and $b = 0.061$. One sees that the data are consistent with an initial linear increase from zero of $1/\Ro_c$ with $1/\Gamma$, with a small quadratic contribution becoming noticeable as $1/\Gamma$ becomes larger.

In order to understand the $\Gamma$ dependence of $1/\Ro_c$,
 we studied the vortex statistics using data obtained from DNS. We used the so-called $Q$-criterion
  \cite{BG86,VE98,VE02b,KCG10}
  to determine the percentage $\bar A$
of the horizontal area that was covered by vortices. 
Using this criterion, implies that 
the quantity  
$Q_{2D}$ \cite{SZCAL09,wil84}, which is a quadratic form of various velocity gradients, was calculated in a plane of 
fixed height.
An area is then identified as ``vortex'' when $Q_{2D}<-0.1\langle |Q_{2D}|\rangle_{v}$, 
were $\langle |Q_{2D}|\rangle_{v}$ is the volume-averaged value of the absolute values of $Q_{2D}$ \cite{SZCAL09}. 
The result of this procedure is shown for different $1/\Ro$ in Fig.\ \ref{WSCLA_fig4} for $\Pra=6.26$.
In Fig.\ \ref{fig:vortices} we plot $\bar A$ as a function of $1/\Ro$ at the edge of the kinetic BL 
(which depends on Ro, see \cite{SCL10})
and at the fixed distance $0.023L$ (the kinetic BL thickness without rotation) from the plates. 
Although there is quite a bit of scatter, the data are consistent with a linear increase of ${\bar{A}}$ for $1/\Ro> 1/\Ro_c$, with a small constant background $\bar A = A_0$  below $1/\Ro_c$. 
The azimuthally averaged vortex density $\langle A \rangle_{\phi}$  
  is given in  Fig.\ \ref{fig:radial}. It shows
that 
the Ekman vortices are inhomogeneously distributed: While in the bulk their fraction  is roughly
constant -- the variations are (presumably) due to insufficient statistics -- 
there are almost no vortices at all close to the side wall, signaling a strong boundary effect.

\begin{figure}
\includegraphics[width=0.39\textwidth]{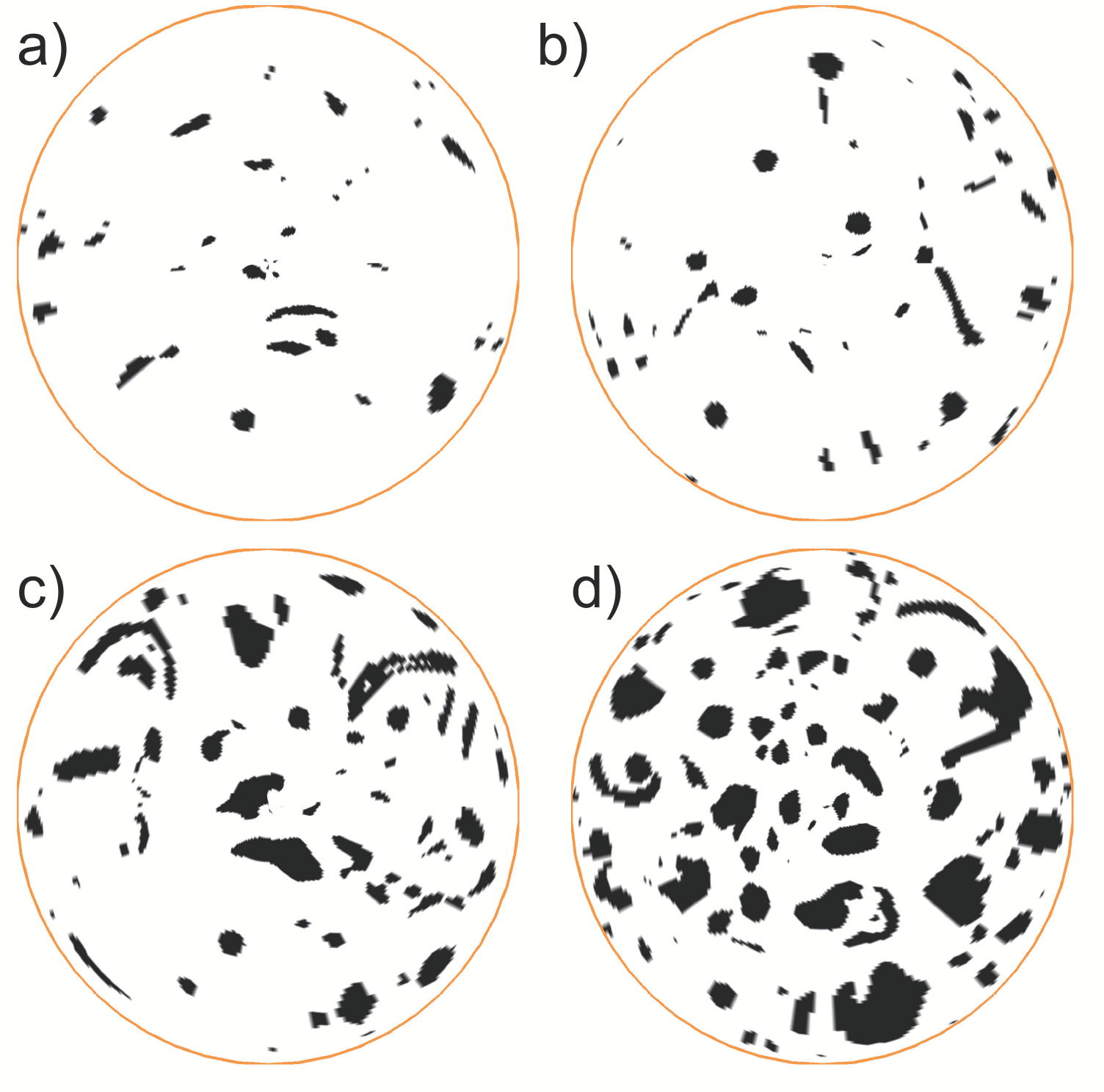}
\caption{The vortices as identified by the $Q_{2D}$ criterion \cite{wil84,SZCAL09} for $\Ra = 2.73\times 10^8$, $\Pra = 6.26$, and $\Gamma = 1$. a) $1/\Ro=2/3$, b) $1/\Ro=1$, c) $1/\Ro=1.54$, and d) $1/\Ro=3.33$. The vortex area increases with increasing $1/\Ro$.
 This trend is quantified  in Fig.\ \ref{fig:vortices}.
}
\label{WSCLA_fig4}
\end{figure}

\begin{figure}
\centering
\includegraphics[width=0.38\textwidth]{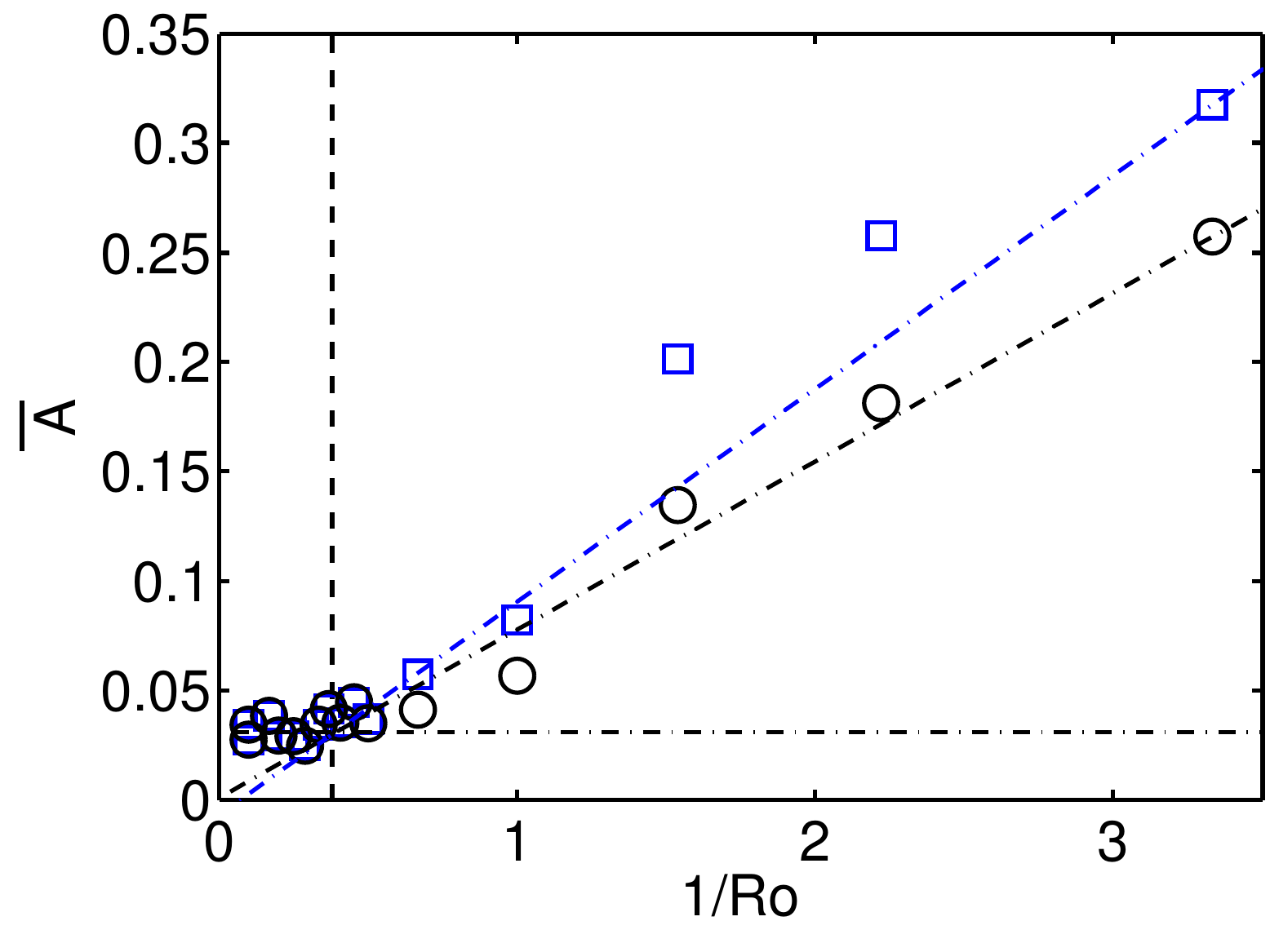}
\caption{Fraction ${\bar{A}}$ of a horizontal slice covered with vortices at the edge of the kinetic BL (circles) and at a distance $0.023 L$ (the kinetic BL thickness without rotation, squares) from the plates as a function of 1/\Ro\ for $\Ra=2.73 \times 10^8$, $\Pra=6.26$, and $\Gamma=1$. The vertical dashed line indicates the bifurcation point at $1/\Ro_c$ and the horizontal dash-dotted line is a background vorticity level $\bar A_0$ present even below $1/\Ro_c$.
}
\label{fig:vortices}
\end{figure}

\begin{figure}
\centering
\includegraphics[width=0.38\textwidth]{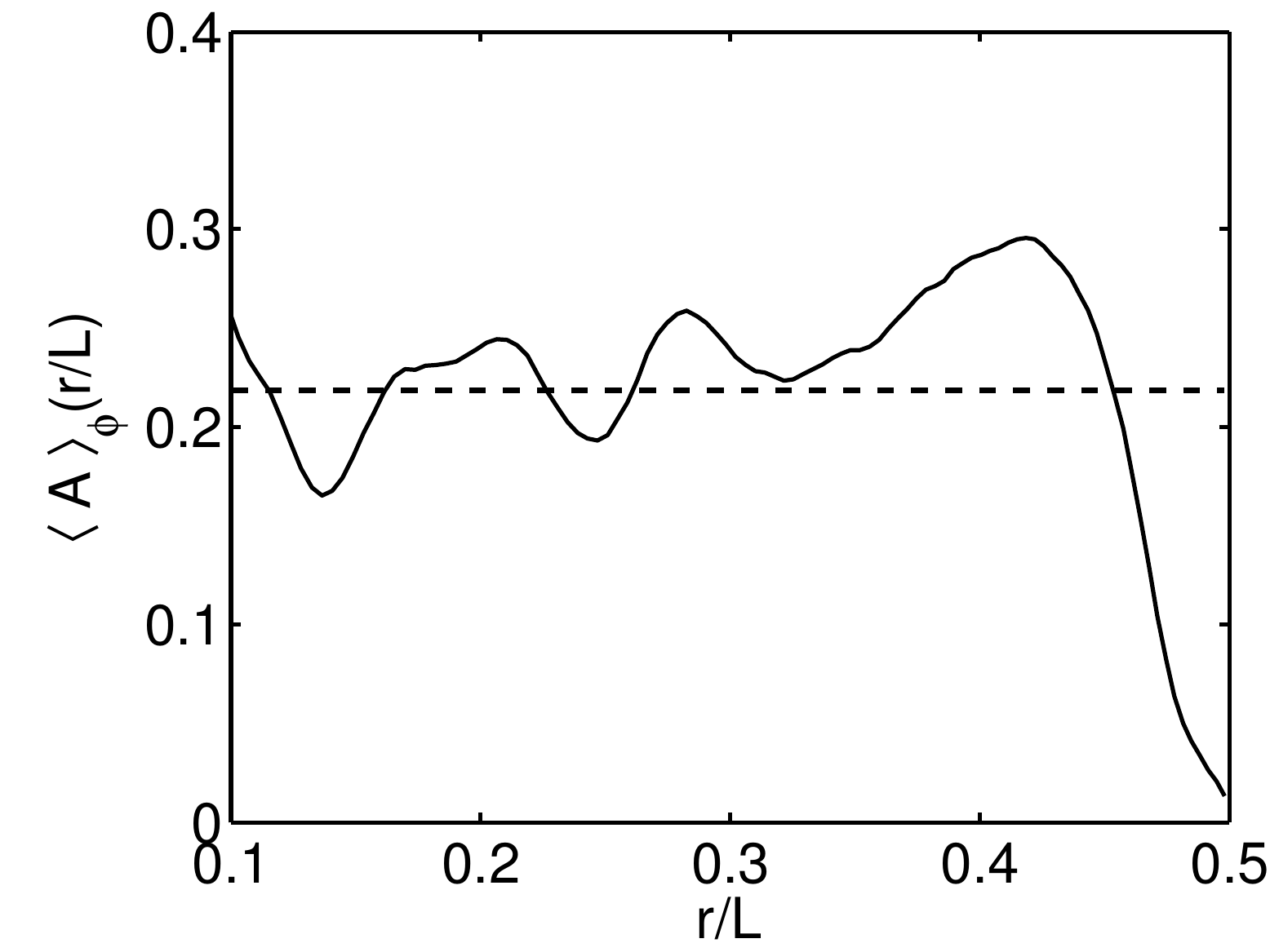}
\caption{Azimuthal average $\langle A \rangle_\phi(r/L)$ of the vortex density $A$ for $\Ra=2.73\times10^8$, $\Pra=6.26$, $\Gamma = 1$, and $2.22<1/\Ro<3.33$. In total, the statistics is based on 8 snapshots.
The dashed line is the uniform case. The density approaches zero close to the side wall. 
}
\label{fig:radial}
\end{figure}

In an effort to understand the existence of a finite onset (see Fig.~\ref{fig:Nu}) of the Ekman-vortex formation and the dependence of the critical inverse Rossby number on $\Gamma$ (see Fig.~\ref{fig:Ro_c}), to elucidate the linear rise and initial slope of $\Nu(1/\Ro)$ above onset (see Fig.~\ref{fig:Nu}), and to explain the rapid decrease of $\langle A \rangle_\phi(r/L)$ near the wall (see Fig.~\ref{fig:radial}), we propose a phenomenological Ginzburg-Landau like model for the local vortex 
density
\be
\dot A = (1/\Ro^2) A - g A^3 + \xi_0^2 \nabla^2 A\ .
\label{eq:GL}
\ee
Here $\dot A$ is the time derivative of $A$. We chose the coefficient of the linear term as $1/\Ro^2$ because for the time independent infinitely extended spatially uniform system it yields a stable solution $A = g^{-1/2}(1/\Ro)$ which implies a vortex density proportional to the rotation rate. 
The term with $\nabla^2A$  represents the lowest-order term of a gradient expansion since terms proportional to $\nabla A$ would lead to an unphysical propagating mode. 

When spatial variations are allowed, the ground state $A = 0$ can be shown to be stable (i.e. to have a growth rate $\sigma < 0$) to disturbances with wave vector $k$ when $1/\Ro$ falls below a neutral curve given by 
\be
1/\Ro_0(k) = \xi_0 k\ .
\ee 
For the finite system it is necessary to introduce appropriate boundary conditions. Here we shall consider a one-dimensional system over the range $-\Gamma/2 \leq x \leq \Gamma/2$ for simplicity and illustrative purposes. The  two-dimensional system with circular boundaries and no azimuthal variation was treated in detail in Ref.~\cite{ACHS81} and yields the same result for $1/\Ro_c$. Since there can be no vortices at the side wall of the sample (see Fig. \ref{fig:radial} where we verified this based on the numerical data), we chose $A(-\Gamma/2) = A(\Gamma/2) = 0$. For the wave number $k_0$ of the lowest mode this yields $k_0 = \pi/\Gamma$.  This in turn gives 
\begin{eqnarray}
\frac{1}{\Ro_c} \equiv  \frac{1}{\Ro_0(k_0)}  =  \frac{\pi\xi_0}{\Gamma} .
\label{eq:Ro_c}
\end{eqnarray}
Thus, consistent with the data in Fig.~\ref{fig:Ro_c}, the model yields the proportionality between  $1/\Ro_c$ and $1/\Gamma$.   We note that the curvature indicated by the quadratic contribution to the fit Eq.~\ref{eq:polynomial} can be accommodated easily by higher-order gradient terms in Eq.~\ref{eq:GL}. 
Comparison with experiment (see Eq.~\ref{eq:polynomial}) gives $\xi_0 = a/\pi = 0.121$. 

To elucidate the rapid decrease of  $\langle A \rangle_\phi(r/L)$ in Fig.~\ref{fig:radial} near $r/L=0.5$, we consider Eq.~\ref{eq:GL} for a semi-infinite system over the range $-\infty < x \leq 0.5$ with the boundary condition $A(x=0.5) = 0$. It yields the solution 
\be
A(x) = (\Ro^2 g)^{-1/2}\tanh{((0.5-x)/\xi)}
\ee
 with 
 \be
 \xi = \sqrt{2}\xi_0 \Ro\ .
 \label{eq:xi}
 \ee
Thus, near the boundaries, the model predicts that the amplitude $A(x)$ of the one-dimensional model, and thus to a good approximation also the azimuthal average $\langle A \rangle_\phi(r/L)$ in Fig.~\ref{fig:radial}, should ``heal" to its bulk value over a length $\xi$. Using a representative $\Ro \simeq 0.4$ for Fig.~\ref{fig:radial}, we estimate from Eq.~\ref{eq:xi} that $\xi \simeq 0.07$. This is roughly consistent with the rapid variation of  $\langle A \rangle_\phi(r/L)$ near $r/L = 0.5$ seen in Fig.~\ref{fig:radial}.

Above the bifurcation the model yields \cite{ACHS81}
\be
{\bar {A}} = \tilde g^{-1/2}\Big ( \frac{1}{\Ro} - \frac{1}{\Ro_c}\Big )~.
\label{eq:A}
\ee
Thus
\be
\frac{\delta \Nu}{\Nu(0)} = S_1(\Gamma)\Big ( \frac{1}{\Ro} - \frac{1}{\Ro_c}\Big )
\label{eq:dNu}
\ee 
which is consistent with the data in Fig.~\ref{fig:Nu}. 
Numerical solutions of the amplitude equation of Ref.~\cite{ACHS81} have shown that the re-normalized coefficient $\tilde g$ in Eq.~\ref{eq:A} is larger than $g$ in Eq.~\ref{eq:GL}. Thus, the initial slope of ${\bar{A}}$ above $1/\Ro_c$ and $S_1(\Gamma)$ in Eq.~\ref{eq:dNu} are reduced by the finite size of the system. The decrease of $S_1(\Gamma)$ with decreasing $\Gamma$ that can be seen in Fig.~\ref{fig:Nu} is also consistent with the model. 

At constant $\Gamma$, $S_1$ depends slightly on the Prandtl number. This suggests that the nonlinear coefficient $g$ in Eq.~\ref{eq:GL}, and thus $\tilde g$ in Eq.~\ref{eq:A},  is dependent on $\Pra$. Similarly, the bifurcation point $1/\Ro_c$ depends slightly on \Pra. This is accommodated in the model Eq.~\ref{eq:GL} by a slightly Prandtl-dependent length scale $\xi_0$.
 
 It remains to be seen whether the phenomena reported and explained here in terms of a finite-size effect have analogies in bifurcations between turbulent states in other systems \cite{rav04,mon07,rav08}.
The more general lesson which is learned is that the Ginzburg-Landau approach, which has been so 
versatile to understand the spatio-temporal dynamics of patterns, can also be useful in understanding
the remarkable bifurcations between turbulent states. 

{\it Acknowledgements:}
We thank Jim Overkamp  for contributing to the experiments with $\Gamma=2$, and Gerald Oerlemans, Chao Sun,
 and Freek van Uittert for the design and construction of the experimental setup in Eindhoven. 
The work of S.W., J.-Q. Z.,  and G.A. was supported by the U.S. National Science Foundation through Grant DMR07-02111. We thank the DEISA Consortium (www.deisa.eu), co-funded through the EU FP6 project RI-031513 and the FP7 project RI-222919, for support within the DEISA Extreme Computing Initiative. The simulations were performed on the Huygens cluster (SARA) and the support from Wim Rijks (SARA) is gratefully acknowledged. RJAMS  was financially supported by the  Foundation for Fundamental Research on Matter (FOM). 


\end{document}